\documentstyle[epsfig,aps,prl,multicol]{revtex}
\draft
\newcommand{\br}{{\bf r}}

\newcommand{\bR}{{\bf R}}
\newcommand{\bk}{{\bf k}}
\newcommand{\bu}{{\bf u}}
\newcommand{\cH}{{\cal H}}

\newcommand{\grad}{{\bf {\nabla}}}

\begin{document}
\title{Theory of plastic vortex creep}

\author{J. Kierfeld, H. Nordborg, and V.M. Vinokur}
\address{Argonne National Laboratory,
Materials Science Division,
9700 South Cass Avenue,
Argonne, Illinois 60439}
\date{\today}
\maketitle

\begin{abstract}
We develop a theory for plastic flux creep in a topologically 
disordered vortex solid phase in type-II superconductors.
We propose a detailed description of the plastic vortex 
creep of the dislocated, amorphous vortex glass in terms 
of  motion of {\em dislocations } driven by a transport current 
$j$. The  {\em plastic barriers} $U_{pl}(j)\propto j^{-\mu}$ 
show power-law divergence at small drives with exponents $\mu=1$ for 
single dislocation creep and  $\mu = 2/5$ for creep of  dislocation bundles.
The suppression of the creep rate is a hallmark of the transition 
from the topologically ordered vortex lattice  to an amorphous 
vortex  glass, reflecting a jump in 
$\mu$ from $\mu = 2/11$, characterizing creep in the 
topologically ordered vortex lattice near the transition, to its plastic 
values.  The lower creep  rates  explain the observed increase 
in apparent critical currents in the dislocated vortex glass.

\end{abstract}

\pacs{PACS numbers: 74.60.Ge, 74.60.Jg, 61.72.Bb}

\begin{multicols}{2}
\textwidth42.5pc
\hsize20.7pc

One of the most fascinating dynamic phenomena of complex systems with
internal degrees of freedom is the thermally activated motion of elastic
media in a random environment (creep) characterized by a highly nonlinear
response to a dc driving force, $F$: $v \sim
\exp{(-\mbox{const}/TF^{\mu})}$, where $v$ is the velocity, $T$ is the
temperature, and $\mu$ is the exponent depending on the geometry and the
dimensionality of the driven medium. The concept of thermally activated
creep,
 which was originally introduced to describe the driven motion of elastic
manifolds through quenched disorder in the context of the dynamics of
dislocation and/or domain walls in  inhomogeneous environments\cite{IV87},
was extended later to driven vortex lattices \cite{FGLV89,F89,Na90}, and has
eventually become a paradigm for the thermally activated dynamics of
disordered media. The concept of creep proved to be especially
successful in describing a wealth of the low temperature transport and
relaxational properties of the vortex state of high temperature
superconductors\cite{blatter}. Recent observations\cite{L98} on 2D
magnets confirmed that domain walls also exhibit creep behavior.
Creep dynamics results from
the scaling of energy barriers $U(j) \sim j^{-\mu}$
which control thermally activated motion. The derivation of such
energy barriers 
diverging at small currents (driving forces)  was based on the
{\it elastic} behavior of the pinned structures; thus 
in the common view 
creep behavior is implicitly attributed to the elastic medium  free of
topological defects.

The description of thermally activated dynamics of  amorphous 
structures containing a large amount of topological defects
remains a major challenge. This is a long standing problem in the
theory of  work
hardening and  related relaxation processes in dislocated solids. In the
context of vortex physics the quest for the description of thermally
activated dynamics of a topologically disordered medium was motivated by
the observation of the disorder-induced transition between a low-field
quasilattice or Bragg glass (BrG)  \cite{Na90,GL}, 
 the phase free of topological defects,  
and a high-field phase characterized by an  enhanced apparent critical
current \cite{Kop}. 
The latter phase was suggested to be a {\it
topologically disordered} vortex state. 
It was shown that the 
 low-field BrG phase is  indeed  {\em stable} against 
dislocation formation \cite{KNH97,DF97}. 
At high magnetic fields the vortex solid undergoes 
a  structural transition  which is described
\cite{Linde,VV98,JK98} as a topological transition between the BrG 
and a dislocated,   amorphous VG (AVG) \cite{KNH97,DF97}.
In a recent series of 
experiments \cite{Ko99,vB00} the phase coexistence characteristic for a 
first-order transition was established, and
creep barriers in the high-field vortex state were shown to 
 diverge {\it faster} than creep
barriers in the low-field elastic phase.

These recent experimental findings  call urgently  for a theoretical
description of the AVG, and the main theoretical problem is to find an 
appropriate 
quantity enabling parameterization of the amorphous phase.
A  first step towards such a quantitative description was
made in  Ref.\cite{KV99} where 
all phase transitions between vortex lattice phases 
were described 
in terms of {\em dislocation-mediated} behavior, and  a 
free energy  functional ${\cal F}[n_{D}]$ for an ensemble of 
directed dislocations
in the presence of thermal fluctuations {\em and} quenched disorder
was  derived ($n_{D}$ is the areal dislocation density).
The BrG-AVG transition was found to be of weak first order 
in accordance with the experimental results of Ref.~\cite{vB00}:
at the BrG-AVG transition dislocations enter with a 
density $n_D \simeq R_a^{-2}$ given by the 
 positional correlation length $R_a$ on which
typical vortex displacements are of the order of  the lattice spacing 
 $a$ \cite{blatter}.
Upon increasing the magnetic field up to the critical point the
dislocation density of the AVG increases to  vortex-liquid-like values 
$n_D \simeq a^{-2}$ such that the AVG and vortex liquid phases become
thermodynamically indistinguishable at the critical point.

In this Letter, building on the aforementioned ideas,
we propose a quantitative description of 
plastic creep in terms of the dislocation degrees of freedom.
We establish and find a critical plastic current $j_{pl}$ 
below which dislocations are {\em collectively pinned} and 
 plastic creep occurs  via the {\em activated} motion
of collectively  pinned dislocation lines.
The critical plastic current is lower than the critical current 
for vortex depinning $j_{pl}<j_c$, hence plastic motion of depinned 
dislocations sets in before viscous flow of the entire vortex lattice can 
occur. 
We calculate the associated {\em plastic} plastic creep energy barriers 
$U_{pl}(j)\sim j^{-\mu_{pl}}$ diverging
{\em infinitely} at $j \to 0$.
We treat  dislocations in the pinned vortex lattice
as  directed elastic strings subject to a 
  pinning force which we derive 
from the Peach-K{\"o}hler force exerted on vortices by the pinning
centers. We show that
an external current sent through a dislocated vortex
lattice generates a   Peach-K{\"o}hler force with a driving component
for dislocation glide.
Knowledge of the pinning and driving forces on the elastic 
dislocation string 
 enables us finally to study the glassy dynamics of the
dislocation, in particular the depinning threshold for dislocation 
glide and the energy barriers for plastic creep below the
depinning threshold.

The energy of a single {\em straight} dislocation 
of length $L$ and with Burger's vector ${\bf b}$
in the FLL
 consists of the core energy  and of the
logarithmically diverging contribution from the long-range elastic
strains \cite{hirth}:
$E_{0} = L E_D ( c_{D} + \ln{\left(L_{\perp}/a\right)} )$,
where $L_{\perp}$ is the lateral system size,
$E_D =  K b^2/4\pi$, $K=\sqrt{c_{44}c_{66}}$ is the {\em isotropized elastic
constant} in the rescaled coordinate 
$z =\tilde{z} \frac{1}{2}\sqrt{c_{44}/c_{66}}$, and $c_D \approx 1$ is 
found numerically ($c_{44}$ and $c_{66}$ are the tilt and shear
moduli of the vortex lattice, respectively). 
Bending of the dislocation line costs an elastic energy associated
with its stiffness $\epsilon_D$.
Hence, the single directed dislocation line -- parameterized by its 
displacement field $\bu_D(z)$ --  is described by the Hamiltonian 
\begin{equation}
\cH_{D}[\bu_D] =  E_0 +  \int dz  
  \frac{1}{2} \epsilon_D (\partial_z \bu_D)^2  
 \label{HD}
\end{equation}
where the stiffness
$\epsilon_D \approx E_D \ln{\left(1/k_za\right)}$
has a logarithmic dispersion due to  the long-range strain field.

Let us now consider the driving force on an  edge  dislocation with 
$b||x$ if 
a transport current $j||y$ is sent through the sample.
Due to Maxwell's equation $\grad \times B = \frac{4\pi}{c} {\bf j}$ 
 the driving current creates a magnetization gradient 
which, in turn, gives rise to shear strains in the vortex lattice:
$\partial_x u_y = \partial_x a = a \frac{2\pi}{c}\frac{j}{B}$. 
Resulting shear stresses give rise to a glide-component of the driving
Peach-K{\"o}hler force\cite{hirth} 
\begin{equation}
 F^{drive}_x = \sigma_{yx} b = 
   b K a \frac{2\pi}{c} \frac{j}{B}
 \label{FPK}
\end{equation}
per length 
(compression stress can only give rise
to dislocation climb which can be neglected as 
 slow process requiring diffusion of
interstitials \cite{NM90}).

The displacements induced by the magnetization gradient can only be
accommodated by the creation of a {\em stationary}  superstructure  of 
regularly spaced dislocation
bands with Burger's vectors with an $y$-component\cite{Braun96} 
in which our  single ``test''-dislocation is moving.
 The superstructure is 
 similar to the grain boundaries formed in an bent  atomic
crystal\cite{hirth}. 
Because such  dislocation bands are essentially free of shear
 stresses\cite{hirth} our  test-dislocation 
 experiences the purely elastic driving force (\ref{FPK}) everywhere 
in between the bands  such that the presence of the superstructure 
does not affect the glide motion.

\begin{figure}[h]
\epsfxsize=3.25in
\epsfysize=2.in
\epsffile{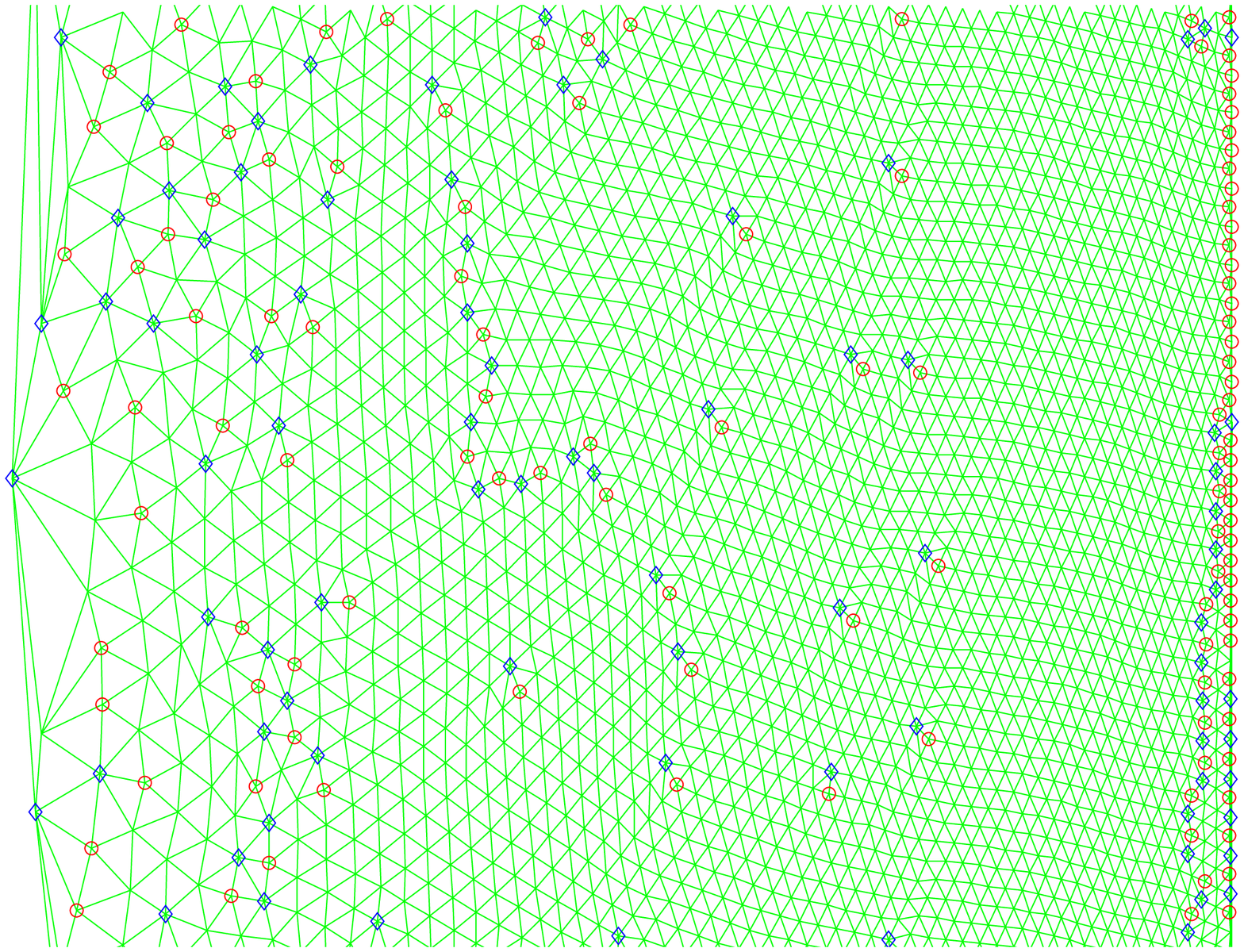}\\
{FIG.~1.~Dislocation bands obtained from numerical integration 
 of the  time-dependent Ginzburg-Landau equation with current
 $j||y$.}
\end{figure}

In the presence 
of  a random pinning potential $V_{pin}(\br)$ in the 
vortex array also the dislocation experiences 
Peach-K{\"o}hler-type pinning forces.
To calculate these we have to 
 determine the  random stress exerted by the pinning 
potential in a frozen-in elastic displacement configuration 
 $u_{el}(\bR,z)$ of the vortex lattice:
$V_{pin}(\bR+\bu_{el},z) = \sigma^{pin}_{ij}(\bR+\bu_{el},z) \grad_i
u_{el,j}$.
The distribution of the pinning stresses
is given by the quenched distribution  of the elastic displacements 
  $u_{el}$. 
This distribution  is determined by 
 the 
different spatial scaling regimes of the 
{\em dislocation-free} collectively pinned vortex array:
(i)
Small distances where vortex displacements $u$ are 
smaller than the coherence length $\xi$ 
 and perturbation theory applies \cite{LO}.
(ii)
Intermediate scales where $\xi \lesssim u \lesssim a$ and
disorder potentials seen by different FLs are effectively {\em
uncorrelated}.  This regime is
captured in so-called {\em random manifold} (RM) models \cite{blatter,GL},
leading to a roughness
$\tilde{G}(\br)=\overline{\langle (\bu_{el}(\br)-\bu_{el}(0))^2 \rangle}
\simeq a^2 (r/R_a)^{2\zeta_{RM}}$
where $\zeta_{RM} \approx 1/5$ for the $d=3$
dimensional RM with two displacement components.
The crossover scale to the asymptotic behaviour is the
{\em positional correlation length} $R_a$ where
the average displacement is of the order of the FL spacing:
$u \approx a$.
(iii)
The asymptotic {\em Bragg glass}
regime where the $a$-periodicity of the FL array becomes important
for the coupling to the disorder and the array is effectively
subject to a {\em periodic} pinning potential with period $a$ \cite{Na90}.
Here the {\em logarithmic} roughness
$\tilde{G}(\br) \approx (a/\pi)^2 \ln{\left(e r/R_a\right)}$,
i.e., $\zeta_{BrG} = {\cal O}({\rm log})$ \cite{Na90,GL}  takes over.

For the physics of dislocations on scales $>a$ only the RM and BrG 
regimes are relevant.
We obtain approximately Gaussian distributed 
quenched stresses with $\overline{\sigma^{pin}_{ij}}=0$ and 
$\overline{\sigma^{pin}_{ij}(\bk)\sigma^{pin}_{ij}(\bk')} = \Sigma^{pin}(k) 
 (2\pi)^3 \delta(\bk+\bk')$ with  $\Sigma^{pin}(k)  = 
  K^2 k^2 G(k)$, i.e., 
\begin{equation}
 \Sigma^{pin}(k)  = 
  K^2a^2 k^{-1} \left\{
\begin{array}{ll}
\mbox{BrG:}
& 1 \\
\mbox{RM:}
& B_{RM} (k R_a)^{-2\zeta_{RM}}
\end{array} \right.
\label{Sigma}
\end{equation}
determined by the elastic correlations $G(k)$ with a numerical 
constant $B_{RM}$.
The RM-result holds for $kR_a >1$ the BrG-result for  $kR_a <1$.

To derive the correct  Peach-K{\"o}hler pinning force on the dislocation 
it is  crucial not only to consider the ``direct''
 quenched pinning stresses
$\sigma_{ij}(\br)$ but also the elastic stresses $\sigma^{el}_{ij}$ 
themselves which are responding to the same  pinning
potential and hence  tend to {\em relax} (longitudinal) components 
of the stress. A lengthy calculation shows that
 if both components are properly added
the pinning Peach-K{\"o}hler force on a dislocation element $d\bR$
\begin{equation}
  d F^{pin}_{\alpha} = \epsilon_{\alpha\beta l} 
  (\sigma^{pin}_{\beta k} +\sigma^{el}_{\beta k}) b_k dR_l
\end{equation}
is rotation-free $(\nabla \times d{\bf F}^{pin})_{\gamma}=0$ such that 
its potential -- the {\em pinning Hamiltonian} -- can be defined:
\begin{eqnarray}
\cH^{pin}_D[\bu_D] &=& \int dz \frac{d\bu_{D,l}}{dz} 
    b_{k} g_{kl}^{pin}(\bu_D,z) 
   \nonumber\\
 \overline{g^{pin}_{kl}(\bk)g^{pin}_{k'l'} (-\bk)}  
   &\simeq& k^{-2} \Sigma^{pin}(k)\delta_{kk'}\delta_{ll'}  =
     K^2 G(k)\delta_{kk'} \delta_{ll'}
\label{Hpin}
\end{eqnarray}
The free energy 
 $\cH_D[\bu_D]+\cH^{pin}_D[\bu_D]-\int dz {\bf F}^{drive}\cdot \bu_D$ 
from (\ref{HD}), (\ref{Hpin}), and (\ref{FPK}) finally 
provides an adequate description for the  study of an ensemble of
pinned dislocation lines.

We start with statics and discuss roughening of the
dislocation line in the presence of disorder. 
The typical pinning energy fluctuations upon 
displacing a dislocation segment $L$ 
over a distance $u_D$ (in the $xz$ glide plane)
 can be calculated from (\ref{Hpin})
\begin{eqnarray}
 E_{pin}^2(L,u_D)  
  &\simeq&  
   b^2 K^2 L u_D \int\limits_0^{L}\!\!\!dz 
  \int\limits_0^{u_D}\!\!\!dx  \int\!\!\! \frac{d^3k}{(2\pi)^3}
    k^2 G(k) e^{ik_z z+ i k_x x}  
\nonumber\\
   &\simeq&  E_D^2   L u_D
     \left\{ \begin{array}{ll}
        \mbox{RM:} & 
         \left(u_D/R_a\right)^{2\zeta_{RM}}  
         \\
        \mbox{BrG:} & 1  
      \end{array}
    \right.
\label{Epin}
\end{eqnarray}
whereas the corresponding 
elastic bending energy of the dislocation is 
$E_{el}(L,u_D) \simeq \frac{1}{2} E_D \ln{(L/a)} u_D^2/L$ (\ref{HD}). 
Optimization gives a dislocation roughness
\begin{equation}
 u_{D}(L)  \simeq  L  \left\{ \begin{array}{ll}
        \mbox{RM:} &   
         \left(L/R_a\right)^{2\zeta_{RM}/(3-2\zeta_{RM})}
                  {\cal O}({\rm log}) 
         \\
        \mbox{BrG:} &  \ln^{-2/3}(L/a) 
      \end{array}
    \right.
\label{rough}
\end{equation}
i.e., exponents  $\zeta_{D} \approx \frac{15}{13}$ for 
RM scaling ($L<R_a$) and $\zeta_{D} \approx 1-log^{2/3}$ for 
BrG scaling.
An instability with respect to dislocation proliferation is signaled 
by anomalous energy gains if $\zeta_D>1$, i.e., in the RM-regime. In the
BrG-regime the energy balance is more subtle and to conclude 
$\zeta_D<1$ or stability one can convert the result (\ref{rough}) into an
approximate renormalization (RG) scheme: the energy gain due to 
roughening is $\Delta E \sim   E_D L  \ln^{-1/3}{(L/a)}
\sim   E_D L  \tilde{\epsilon}_D(L)^{-1/3}$,
 where 
the logarithmic correction is identical to the dimensionless 
line tension $\tilde{\epsilon}_D = 
\epsilon_D/E_D$ on the scale $L$. 
 Interpreting 
$\Delta E/(E_D L)$ as disorder correction to the 
line tension $\tilde{\epsilon}_D(L)$  on the scale $L$ and summing
 these corrections successively on each scale,
together with the bare tension 
$\tilde{\epsilon}_D^0(L) = \ln{(L/a)}$,
 one obtains an integral RG equation 
\begin{equation}
 \tilde{\epsilon}_D(\ln{L}) = 
    \int_0^{\ln{L}} 
     d\ell (1 \pm  \tilde{\epsilon}_D(\ell)^{-1/3} ) 
\label{RG}
\end{equation}
equivalent  to the result  of  Ref.~\cite{DF97}. Integration 
shows  that  corrections to $\tilde{\epsilon}_D^0(L)$ are irrelevant
and hence  the stability of the BrG regime with
respect to dislocation formation\cite{DF97}.
The detailed stability analysis for both regimes 
was given in Ref.~\cite{KV99}: the 
BrG-AVG transition is weakly first-order, and 
dislocations proliferate with the density $n_d \simeq R_a^2$ 
defined by the  scale $R_a$ of crossover
between the unstable RM and the stable BrG regimes.

Now we extend our scaling analysis to 
 the dynamic behaviour of the driven dislocation
line in quenched disorder. 
Minima of the pinning potential are typically  distances 
$u_D \sim a$,  given by the core size of the dislocation, apart. 
A dislocation confined within {\em one} minimum $u_D \le a$ 
is collectively pinned on a {\em plastic pinning length}
\begin{equation}
  L_{pl} \simeq a \left(\frac{R_a}{a}\right)^{2\zeta_{RM}/3} < R_a  
\label{LcD}
\end{equation}
which is the analogue of the Larkin pinning length $L_c \simeq \xi
\delta^{-1/3}$ 
of the single vortex  where $\delta$ is the dimensionless pinning 
 strength of Ref.~\cite{blatter}. 
To depin the dislocation the driving force (\ref{FPK})
 has to exceed the pinning force $F^{pin}(L_{pl}) 
\simeq \frac{E_D}{a} L_{pl} \left(a/R_a\right)^{4\zeta_{RM}/3}$
on  a segment of length $L_{pl}$.
This determines a {\em  critical plastic current} $j_{pl}$ 
\begin{equation}
\!\! j_{pl} \simeq \frac{c}{8\pi^2}  \frac{B}{a}  
   \left(\frac{a}{R_a}\right)^{4\zeta_{RM}/3}
 \simeq  j_0 \left(\frac{2\pi H_{c2}}{B}\right)^{-7/4} \delta^{5/18}
\label{jpl}
\end{equation} 
where  $j_0 \simeq \frac{c}{6\sqrt{3} \pi} H_{c2}/\xi$ is the
depairing current.
Comparing this result to the depinning 
current $j_c \simeq j_0 \delta^{2/3}$ for a single vortex, one
finds $j_{pl}< j_c$ for typical disorder strengths 
 $\delta \sim 10^{-3}$. This shows that plastic motion of dislocations
will set in  even if the vortices themselves are still pinned. Thus plastic
motion is the dominating transport mechanism in the dislocated
AVG phase. 

However, the existence of a current $j_{pl}>0$ also
demonstrates that dislocations can be {\em pinned} at low 
currents and plastic motion for $j\ll j_{pl}$ will only occur 
through activation over {\em diverging} plastic energy barriers 
$U_{pl}(j)\sim j^{-\mu_{pl}}$ giving rise to plastic creep. 
The typical segment size $L(j)$  
for activated motion at  $j\ll j_{pl}$
is determined by balancing the energy gain by the driving force  
$F^{drive} L^{1+\zeta_D}$
against the 
pinning energy $E_{pin}(L,u_D(L))\simeq  E_D L^{2\zeta_D-1}$ 
of the rough dislocation line. 
This yields 
\begin{equation}
U_{pl}(j) \simeq  E_D a \left(\frac{a}{R_a}\right)^{2\zeta_{RM}/3}
     \left(\frac{j_{pl}}{j} \right)^{(2\zeta_D-1)/(2-\zeta_D)} 
\label{UjD}
\end{equation}
and we obtain the scaling law $\mu_{pl} = (2\zeta_D-1)/(2-\zeta_D)$ 
(as for single vortex creep\cite{blatter})
relating  the  plastic creep exponent $\mu_{pl}$
to the dislocation roughness. We find 
$\mu_{pl} = \frac{17}{11}$ in the RM-regime ($L(j)<R_a$) and 
$\mu_{pl} = 1$ in the BrG-regime ($L(j)>R_a$).
Both exponents are considerably larger than their counterpart
$\mu = 2/11$ for elastic 
single  vortex creep showing  that plastic creep 
rates are much smaller than elastic creep rates.

So far we focused on a {\em single} dislocation. Now we turn to 
interacting dislocations.  On large scales exceeding the dislocation 
spacing  $R_D$, which varies from $R_D \simeq R_a$ at the AVG-BrG 
transition to $R_D \simeq a$ at the critical point \cite{KV99}, 
interactions become essential and plastic creep is governed by motion of 
dislocation {\em bundles} in a glide plane ($xz$-plane).
Deformations $u_D(x,z)$ of such a 
  2D bundle  can be described by an elastic  Hamiltonian
with   tilt modulus 
$K_z \simeq E_D/R_D$ and the compression modulus 
$K_x \simeq  R_D \partial_{R_D}^2 (R_D^2f(R_D))$ 
which can be calculated from the dislocation free 
energy $f(R_D)$, see Ref.~\cite{KV99}
(in the absence of disorder one finds $K_x \simeq E_D/R_D$).
Including the pinning energies we obtain the Hamiltonian
\begin{eqnarray}
 \cH[u_D(x,z)] &=& \int dxdz \frac{1}{2} \left\{ 
     K_x (\partial_x u_{D})^2 +  K_z (\partial_z u_{D})^2 \right\} 
      \nonumber\\
   &&     + \sum_i \int dz \cH^{pin}_D[{\bf b}_i,\bu_D(i R_D,z)]
\label{Hbundle}
\end{eqnarray}
The  dislocation bundle contains  dislocations of {\em opposite} signs
with the same density
to avoid the accumulation of stress. Hence the sum over the
dislocation index $i$ in (\ref{Hbundle}) goes over {\em alternating}
Burger's vectors ${\bf b}_i ||x$. On scales $L_x\gg R_D$ dislocations
couple effectively as dipoles to  disorder and we obtain 
for the bundle disorder energy fluctuations
$E^{b}_{dis}(L_z,L_x,u_{D}) \simeq  
 E_{pin}(L_z,u_D)$
with $E_{pin}(L_z,u_D)$ from  (\ref{Epin}). 
This has to be balanced against the  elastic energy 
$E_{el}(L_z,L_x,u_D) \simeq \sqrt{K_xK_y}  u_D^2$ with 
$L_z \simeq  \sqrt{{K_z}/{K_x}} L_{x}$ resulting in a roughness 
\begin{equation}
    u_{D}(L_z)  \sim L_z^{1/3}R_D^{2/3}  \left\{ \begin{array}{ll}
        \mbox{RM:} &  
   \left(\frac{L_z^{1/3}R_D^{2/3}}{R_a}
              \right)^{\frac{2\zeta_{RM}}{3-2\zeta_{RM}}}
         \\
        \mbox{BrG:} &   1
      \end{array}
    \right. ,
\label{bundlerough}
\end{equation}
i.e., a {\em reduced} 
dislocation bundle roughness $\zeta_{D} \approx \frac{5}{13}$ for 
RM scaling ($L<R_a$) and $\zeta_{D} \approx \frac{1}{3}$ for 
BrG scaling.
It is easy to establish the corresponding plastic creep exponents
for bundle creep  along the same lines as for the single dislocation, 
in particular one finds the same  scaling relation 
$\mu_{pl}= (2\zeta_D+d-2)/(2 - \zeta_D)$ as  for $d$-dimensional 
vortex bundles~\cite{blatter}. 
This gives $\mu_{pl} = \frac{10}{21}$ in the RM-regime  and 
$\mu_{pl} = \frac{2}{5}$ in the BrG-regime.
The crossover from single dislocation to bundle scaling happens 
for currents $j<j^{b}$ where $L(j^{b}) \simeq \sqrt{K_z/K_x} R_D$. 
For $R_D \simeq a$  one finds  $j^{b} \simeq j_{pl}$,  which 
means that only  plastic bundle creep 
can be measured above the
critical point defined by $R_D \simeq a$.
Though we do think that our results apply rather to the YBCO-compound
than the strongly layered BSCCO-compound, 
in Ref.~\cite{Ko99} a creep exponent strikingly close to 
$\mu_{pl} = \frac{2}{5}$ has been measured within the AVG phase.

In the high-field regime above the critical point there is no
melting  phase transition but a crossover in 
the resistive behavior due to  {\em thermal depinning} of dislocations 
will  still be observable.
The single pinned dislocation line thermally depins if
$\langle u_D^2 \rangle_{th}(L_{pl}) \simeq a^2$, i.e., 
when the thermal fluctuations of a segment of the size of the 
 plastic pinning length 
increase beyond a lattice spacing 
 which leads to considerable 
smoothing of the pinning potential. This happens at a 
{\em plastic depinning temperature} $T_{pl} \simeq  
E_D a \left(a/R_a\right)^{2\zeta/3}$. 
Using results of Ref.\cite{KV99}, we find 
$T_{pl} \simeq T_m \simeq 1.6 E_D a$ at the critical point.
this suggests that at the critical point
the first-order melting line terminates and 
transforms into a thermal depinning line $T_{pl}(B) \sim B^{-3/8}$
describing the  dynamic crossover to reduced depinning currents
and  thermal activation over all plastic  barriers $U_{pl}(j)<T$.

In conclusion, we have developed a theory of plastic creep 
in terms of the actual dislocation dynamics in the pinned 
vortex lattice. We find infinite barriers for plastic 
vortex transport in agreement with  experimentally observed 
low creep rates or apparently high critical currents. The obtained 
results are relevant for and are 
easily extended to other systems where glassy dynamics is 
controlled by topological defects, for example, charge density 
waves in disordered crystals and/or work-hardened solids.

We thank A.E.\ Koshelev  for stimulating
discussions.
This work was supported by Argonne National Laboratory through
U.S.\ Department of Energy, Office of Science under contract
No.~W-31-109-ENG-38, and by NSF-STC under Contract
No.~DMR91-20000~STcS.
J.K.\ acknowledges support from the Deutsche Forschungsgemeinschaft
under Grant No.~KI~662/1.

\end{multicols}

\end{document}